\documentclass[twocolumn,amsmath,amssymb,prb]{revtex4}
\usepackage[dvipdfmx]{graphicx,color}

\makeatletter

\renewcommand\thefigure{\arabic{figure}}
\renewcommand{\thesection}{\arabic{section}}
\renewcommand{\thesubsection}{\thesection.\arabic{subsection}}

\def\p@subsection{}
\def\p@subsubsection{}

\makeatother

\begin{document}

\title{
Heat and Charge Current Fluctuations and the Time-Dependent Coefficient of Performance for a Nanoscale Refrigerator
}
\author{Hiroki Okada and Yasuhiro Utsumi}
\affiliation{Department of Physics Engineering, Faculty of Engineering, Mie University, Tsu, Mie 514-8507, Japan}

\begin{abstract}
We theoretically investigate the coefficient of performance (COP) of a mesoscopic thermoelectric refrigerator realized by using a tunnel junction.
We analyze the influence of particle and heat current fluctuations on the COP out of the equilibrium regime.
We calculate the average COP by using full counting statistics and find that it depends on the measurement time $\tau$.
The deviation from the macroscopic COP value can be expressed with the Skellam distribution at all times.
This result enables us to improve the Gaussian approximation valid within the linear response regime, which cannot predict the average COP in the limit of $\tau \to 0$.
We illustrate the time dependence of the average COP and find that in the short-time regime, the average COP possesses a minimum.
In order to confirm the physical consistency far from equilibrium, we propose checking the correlation coefficient between the particle and the heat currents in addition to the positivity of the entropy production rate.
\end{abstract}

\maketitle

\section{Introduction}
The quantum thermodynamics in nanoscale circuits has attracted much attention\cite{Pekola2015}.
In particular, there has been continuous interest in thermoelectric effects.
Three decades ago, the thermoelectric transport theory in multi-terminal quantum conductors was established on the basis of the Landauer--B\"uttiker formula\cite{Sivan1986}.
It has recently been applied to study microscale heat engines, which convert the input charge (particle) current to the output heat current, and vice versa
\cite{Entin-Wohlman2010,Entin-Wohlman2012,Sanchez2011}.
A transducer that converts the heat current into the charge current has been proposed that uses a three-terminal setup~\cite{Entin-Wohlman2010,Entin-Wohlman2012,Sanchez2011}, a molecular bridge strongly coupled to a thermal bath\cite{Entin-Wohlman2010,Entin-Wohlman2012}, or a quantum dot capacitively coupled to another quantum dot acting as a fluctuating gate voltage\cite{Sanchez2011}.
The thermopower and the efficiency in quantum conductors have been discussed~\cite{Crepieux2016,EymeoudPRB2016}.
In the linear response regime, the maximum power and the maximum efficiency have been investigated ~\cite{BenentiPRL2011,ProesmansPRX2016,ProesmansPRL2016}.
Also, the time dependence of the energy or the heat transport in externally driven quantum conductors has been investigated~\cite{MoskaletsPRB2002,ArracheaPRB2007,MoskaletsPRB2009,CrepieuxPRB2011,LudovicoPRB2014,EspositoPRL2015,BruchPRB2016}.
Experimentally, a mesoscopic refrigerator has already been realized by exploiting an SINIS junction\cite{Saira2007}.
Furthermore, the thermoelectric effect has been experimentally\cite{Brantut2012, Brantut2013} and theoretically\cite{Grenier2014} investigated in mesoscopic conductors fabricated by using ultracold atoms.
Recent progress in the quantum thermoelectrics in the nonlinear regime is reviewed in Refs. [\onlinecite{Sanchez2016}] and [\onlinecite{Benenti2013}].

In addition, the discussion of the Onsager symmetry in thermoelectric transport has recently been extended to particles, heat, and spin transport~\cite{JacquodPRB2012,WangPRB2015}.

\begin{figure}[ht]
        \begin{center}
            \includegraphics[width=0.7 \columnwidth]{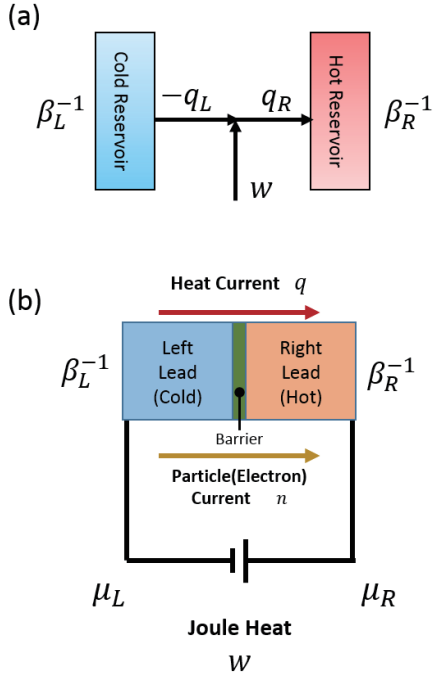}
        \end{center}
        \caption{
(Color online) (a) Schematic picture of the refrigerator.
It consists of hot and cold reservoirs with temperatures $\beta^{-1}_{R}/k_{\rm B}$ and $\beta^{-1}_{L}/k_{\rm B}$, respectively.
The external work $w$ removes heat $-q_{L}$ from the cold reservoir and emits heat $q_{R}=w-q_{L}$ to the hot reservoir.
(b) Mesoscopic refrigerator composed of a tunnel junction.
The right (left) lead corresponds to the hot (cold) reservoir in (a).
The voltage source generates Joule heat, which corresponds to the external work $w$.
}
        \label{fig:setting}
    \end{figure}

In the present paper, we consider a refrigerator composed of a tunnel junction.
Figure~\ref{fig:setting}(a) shows the refrigerator, which consists of hot and cold reservoirs with inverse of temperatures $k_{\rm B}\beta_{R}$ and $k_{\rm B}\beta_{L}$, respectively, where $k_ {\rm B} $ is the Boltzmann constant.
By performing work $w$, it removes heat $-q_{L}$ from the cold reservoir and emits the rest of the heat $q_{R} = w-q_{L}$ to the hot reservoir.
The property of the refrigerator is evaluated by using the coefficient of performance (COP) defined as $\phi_{L} = -q_{L}/w$\cite{Entin-Wohlman2015}.
It is convenient to introduce a symmetrized form, which cannot exceed the following Carnot limit:
    \begin{align}
        \phi \equiv \phi_{L} + \frac{1}{2} =\frac{1}{2}\frac{q_{R}-q_{L}}{q_{R}+q_{L}}
             \leq -\frac{\beta}{\delta \beta} = \phi_{\rm C} ,\label{ineqCarnot}
    \end{align}
where $\delta \beta = \beta_{R}-\beta_{L}<0$ and $\beta = (\beta_{R}+\beta_{L})/2$.
Figure \ref{fig:setting}(b) is a schematic picture of the tunnel junction.
The right and left leads correspond to the hot and cold reservoirs, respectively.
The voltage source generates the chemical potential difference $\delta \mu = \mu_{L}-\mu_{R}>0$ and allows electrons to carry the heat from the left lead to the right lead.
In this thermoelectric refrigerator, the work done by the voltage source $w$ is equal to the Joule heat.

In a macroscopic device, the fluctuation of the COP is negligible.
However, in a nanoscale device, both the heat current and the charge current fluctuate\cite{Seifert2012}.
Therefore, the COP also fluctuates and it is necessary to consider its probability distribution\cite{VerleyNC2014,VerleyPRE2014,Esposito2015,JiangPRL2015,Polettini2015}.
The idea of the probability distribution of efficiency has been introduced and analyzed in detail in the linear response regime in Refs.~[\onlinecite{VerleyNC2014}] and [\onlinecite{VerleyPRE2014}].
Analysis based on a microscopic model has been performed by using the theory of full counting statistics (FCS)\cite{Esposito2015}.
The efficiency statistics of mesoscopic transducers with broken time-reversal symmetry has also been investigated\cite{JiangPRL2015}.
The above studies mainly focused on the long-time limit.
The short-time behavior of the probability distribution of the COP has been analyzed in the linear response regime, where the Gaussian approximation is valid\cite{Polettini2015}.
In the present paper, we will go beyond the Gaussian approximation and analyze the time dependence of the COP in
the framework of FCS~\cite{Esposito2009}.
We will mainly discuss the average COP and demonstrate that it is time-dependent because of particle and heat current fluctuations.

The paper is organized as follows.
In Sect.~\ref{sec:MDL}, we introduce the model Hamiltonian (Sect.~\ref{sec:MHC}).
Then we define the probability distribution of the COP of the refrigerator.
Within the second-order perturbation expansion in the tunnel coupling (Sect.~\ref{sec:SOP}), we provide an analytical formula for the average COP (Sect.~\ref{sec:ACoP}).
We also evaluate the expression for the average COP in the long-time limit (Sect.~\ref{sssec:LTL}) and short-time limit (Sect.~\ref{sssec:STL}) and discuss the fluctuation theorem (Sect.~\ref{sec:FTCoP}).
On the basis of the resulting average COP, we improve the Gaussian approximation in Sect.~\ref{sec:MGA}.
In Sect.~\ref{sec:TECoP}, we illustrate our results.
We also discuss the correlation coefficient between the particle and the heat currents.
Finally, we summarize our results in Sect.~\ref{sec:SMR}.
Technical details are given in appendices.

\section{Probability Distribution of COP\label{sec:MDL}}
\subsection{Model Hamiltonian and cumulant generating function\label{sec:MHC}}

The Hamiltonian of the tunnel junction is written as
    \begin{align}
        \hat{H} =& \hat{H}_{L} + \hat{H}_{R} + \hat{V} . \label{Ham}
    \end{align}
Here $\hat{H}_{r}$ describes the Hamiltonian of lead $r$ ($r=L,R$),
    \begin{align}
		\hat{H}_{r} =& \sum_{\nu} \epsilon_{r \nu} \hat{a}_{r \nu}^{\dagger} \hat{a}_{r \nu} ,
    \end{align}
where $\hat{a}_{r\nu}$ ($\hat{a}_{r\nu}^{\dagger}$) is the operator that annihilates (creates) an electron in energy level $\nu$ of lead $r$.
The tunnel Hamiltonian is given by
	\begin{align}
		\hat{V} =& \sum_{\nu , \nu '} \Omega_{\nu \nu '}(\hat{a}_{R \nu}^{\dagger} \hat{a}_{L \nu'} +\hat{a}_{L \nu'}^{\dagger} \hat{a}_{R \nu}) ,
	\end{align}
where $\Omega_{\nu \nu'}$ is the tunnel matrix element associated with the hopping between level $\nu$ in the right lead and level $\nu'$ in the left lead.

Let us introduce the joint probability distribution of the particle current and heat current.
In this paper, we adopt a two-time measurement protocol\cite{Kurchan2000,Klich2002}.
\begin{itemize}
\item First, we assume that the left and right leads are separated and in equilibrium.
Here, the initial equilibrium density matrix for decoupled leads, which is denoted by $\hat{\rho}_{0}$, is written as
    \begin{align}
        \hat{\rho}_{0} = \prod_{r=L,R} \frac{e^{-\beta_{r} (\hat{H}_{r}-\mu_{r} \hat{N}_{r})}}
                        {{\rm Tr} \left( e^{-\beta_{r} (\hat{H}_{r}-\mu_{r} \hat{N}_{r})} \right)},
    \end{align}
where $\hat{N}_{r}=\sum_{\nu}\hat{a}^{\dagger}_{r\nu}\hat{a}_{r\nu}$ is the operator of the particle number in lead $r$.
\item Then, the first measurement is taken and a many-body state with the Fock representation
  \begin{align}
    |{\bf a} \rangle =& | a_{R1} \cdots a_{R \nu} \cdots a_{L1} \cdots a_{L \nu '}\cdots \rangle
  \end{align}
is chosen.
Here $a_{r\nu}$ is the number of particles occupying level $\nu$ of lead $r$.
\item At time $t_{0}$, the left and right leads are coupled, and the particles begin to tunnel between the leads.
\item The left and right leads are decoupled at time $t\, (>t_{0})$, and then, the second measurement is taken.
We write the final state as $|{\bf b}\rangle$.
\end{itemize}
\noindent
The joint probability to find $|{\bf a}\rangle$ as the initial state at time $t_{0}$ and $|{\bf b}\rangle$ as the state at time $t$ is given by
    \begin{align}
		P({\bf a} , {\bf b}) = \langle {\bf a}|\hat{\rho}_{0}|{\bf a} \rangle |\langle
		{\bf b}|e^{-\frac{i}{\hbar}\hat{H}(t-t_{0})}| {\bf a} \rangle|^{2} \label{P(a,b)} .
    \end{align}
By using Eq.~(\ref{P(a,b)}), we define the joint probability distribution of the number of particles $n$ transferred from the left lead to the right lead, the amount of heat transfer $q$, and the Joule heat $w$ as
	\begin{align}
		 P(n,q,w) =& \sum_{{\bf a},{\bf b}} P({\bf a},{\bf b}) \,\delta_{n,(N_{R}-N_{L})/2} \nonumber \\
		 			\times & \delta(q- (q_{R}-q_{L})/2)\,\,
		 			\delta(w-(q_{R}+q_{L})), \label{P(n,q,w)}
	\end{align}
		where
	\begin{align}
		N_{r} =& \sum_{\nu}(b_{r\nu}-a_{r\nu}) , \\
		q_{r} =& \sum_{\nu}(\epsilon_{r\nu}-\mu_{r})\,(b_{r\nu}-a_{r\nu}).
	\end{align}
By using the joint probability distribution~(\ref{P(n,q,w)}), we can define the probability distribution of the COP of the refrigerator as~\cite{VerleyNC2014,VerleyPRE2014,Esposito2015,JiangPRL2015,Polettini2015}
    \begin{align}
        P(\phi) =& \int_{-\infty}^{\infty} dq \int_{-\infty}^{\infty} dw
        \sum_{n\neq0} P(n,q,w) \delta \left( \phi-\frac{q}{w} \right) \ \ .
        \label{Prob2}
    \end{align}
In the following, we will calculate it from the microscopic Hamiltonian (\ref{Ham}).
Here, it is convenient to calculate the cumulant generating function (CGF), which is the logarithm of the Fourier transformation of the joint probability distribution (\ref{P(n,q,w)}),
    \begin{align}
        W(\lambda,\xi,\Xi) = {\rm ln} \int dq dw \sum_{n}P(n,q,w)e^{in\lambda + i\xi q + i\Xi w}.
        \label{CGF}
    \end{align}
From the derivative of the CGF we can calculate, for example, the averages $\langle n \rangle$ and $\langle q \rangle$, the variances $\sigma_{n}^{2}$ and $\sigma_{q}^{2}$, and the covariance $C_{nq}$ [see Eqs. (\ref{n})-(\ref{Cnq})].

\subsection{Second-order expansion\label{sec:SOP}}

The CGF (\ref{CGF}) can be written as
    \begin{align}
        W(\lambda,\xi,\Xi) =& {\rm ln} {\rm Tr} \left( \hat{\rho}_{0} \hat{U}_{I,-\lambda,-\xi,-\Xi}^{\dagger}\,\hat{U}_{I,\lambda,\xi,\Xi} \right)
        \label{CGF0}.
    \end{align}
The modified time evolution operator is
    \begin{align}
        U_{I,\lambda,\xi,\Xi} =& \hat{T} {\rm exp}\left(- \frac{i}{\hbar} \int_{t_{0}}^{t} dt'
                        e^{ i \hat{\theta} } \, \hat{V}_{I}(t') \,e^{ -i \hat{\theta} } \right) , \\
        \hat{\theta} =& \frac{1}{2}\sum_{r=L,R} \left[ s_{r}\frac{\lambda}{2} \hat{N}_{r}
                        + \left( \Xi + s_{r}\frac{\xi}{2} \right)(\hat{H}_{r}-\mu_{r} \hat{N}_{r}) \right] ,\label{CountField}
    \end{align}
where $s_{R}=1$, and $s_{L}=-1$, and $\hat{T}$ is the time-ordering operator.
The operators with the subscripts $I$ stand for those in the interaction picture,
$\hat{V}_{I}(t)=e^{\frac{i}{\hbar}(\hat{H}_{L}+\hat{H}_{R})(t-t_{0})}\hat{V}e^{-\frac{i}{\hbar}(\hat{H}_{L}+\hat{H}_{R})(t-t_{0})}$
and
$\hat{U}_{I}(t,t_{0})=e^{\frac{i}{\hbar}(\hat{H}_{L}+\hat{H}_{R})(t-t_{0})}\,e^{-\frac{i}{\hbar}\hat{H}(t-t_{0})}$.

For the tunnel junction, we can perform perturbation expansion of the CGF (\ref{CGF0}) up to the second order in $\hat{V}$.
The first-order contribution vanishes after taking the trace.
The lowest nonvanishing contribution becomes
    \begin{align}
        W (\lambda,\xi,\Xi) \simeq& F(\lambda,\xi,\Xi) - F(0,0,0) \ , \label{CGF1}
        \\
        F(\lambda,\xi,\Xi) =& \tau \, G(\xi)\,{\rm cosh} \left(\beta \frac{\delta \mu}{2} +i\lambda + i\,\Xi\, \delta \mu \right)
        \nonumber \\ &+ \tau \,
        \varepsilon(\xi)\,{\rm sinh} \left(\beta \frac{\delta \mu}{2} +i\lambda + i\,\Xi\, \delta \mu \right)
        \label{CGF2}
    \end{align}
if the measurement time $t$ is sufficiently long.
Here we introduce the dimensionless measurement time as
    \begin{align}
        \tau =& (t-t_{0}) \frac{4\pi |\Omega|^{2}}{\hbar}
        \frac{D_{R}(\bar{\mu})D_{L}(\bar{\mu})}{{\beta}} ,
    \end{align}
where $D_{r}(\epsilon)=\sum_{\nu}\delta(\epsilon-\epsilon_{r\nu})$ is the density of states (DOS) of lead $r$ and $\bar{\mu} =
(\mu_{L}+\mu_{R})/2$ is the average chemical potential.
$\tau$ is the product of the measurement time $t-t_{0}$ and the thermal noise, which is the electric conductance in the linear response regime divided by the inverse temperature.

The coefficients $G(\xi)$ and $\varepsilon(\xi)$ [Eqs. (\ref{N1}) and (\ref{N2})] are the dimensionless conductance and electromotive field, respectively.
The averages, the variance, and the covariance [Eqs.~(\ref{n}, \ref{q}, \ref{sigman}, \ref{Cnq})] are derived as
    \begin{align}
        \left(
        \begin{array}{c}
                \langle n \rangle \\ \langle q \rangle
            \end{array}
            \right) =& \tau
            \left(
            \begin{array}{cc}
                G(0) & \varepsilon(0) \\
                \partial_{i\xi}\varepsilon(0) & \partial_{i\xi}G(0)
            \end{array}
            \right)
            \left(
            \begin{array}{c}
                {\rm sinh}\left( \beta\frac{\delta \mu}{2} \right) \\ {\rm cosh}\left( \beta\frac{\delta \mu}{2} \right)
            \end{array}
            \right), \label{Averages} \\
            \left(
            \begin{array}{c}
                \sigma^{2}_{n} \\ C_{nq}
            \end{array}
            \right) =& \tau
            \left(
            \begin{array}{cc}
                G(0) & \varepsilon(0) \\
                \partial_{i\xi}\varepsilon(0) & \partial_{i\xi}G(0)
            \end{array}
            \right)
            \left(
            \begin{array}{c}
                {\rm cosh}\left( \beta\frac{\delta \mu}{2} \right) \\ {\rm sinh}\left( \beta\frac{\delta \mu}{2} \right)
            \end{array}
            \right).\label{Vars}
    \end{align}
In particular, the averages (\ref{Averages}) reproduce the linear response theory upon expanding $G(\xi)$ and $\varepsilon(\xi)$ (Eqs. (\ref{N1}) and (\ref{N2})) in $\beta\, \delta \mu$ and $\delta \beta$ up to the first order
(Appendix~\ref{ssec:IaE}),
    \begin{align}
        \left(
        \begin{array}{c}
            \langle n \rangle \\ \langle q \rangle
        \end{array}
        \right)
        /\left( \frac{\tau}{2} \right)
        \simeq & \,{\bf L}
        \left(  \begin{array}{c}
        {\beta\, \delta \mu} \\ {\delta \beta}
        \end{array} \right) \label{lincum},
        \\
        {\bf L} =& \left( \begin{array}{cc}
                             1 &  D_{rat}\, l_{\beta} \\ D_{rat}\, l_{\beta} &
                             l_{\beta} \end{array}\right) , \label{MtrxLR}
        \\
        l_{\beta} =& \frac{1}{\beta^{2}}\frac{\pi^{2}}{3},
    \end{align}
where $D_{rat} \equiv \partial_{\bar{\mu}}{\rm ln}D_{L}(\bar{\mu})D_{R}(\bar{\mu})$.
Here we remark that the transport matrix (\ref{MtrxLR}) is normalized by the dimensionless time $\tau$, which contains all the system parameters.

\subsection{Average COP\label{sec:ACoP}}

By combining the CGF~(\ref{CGF1}) and the definition of the probability distribution of the COP (\ref{Prob2}), we derive the average of the COP (see Appendix \ref{ssec:jPDF} for details),
    \begin{align}
        \langle \phi \rangle_{\tau} =& \frac{\langle q \rangle}{\delta \mu\, \langle n \rangle}
        +\frac{C(\tau)}{\delta \mu}
        \left( \frac{ \langle q \rangle }{ \langle n \rangle } -\frac{C_{nq}}{\sigma^{2}_{n}} \right)
        \label{aveCoP} ,
        \\
        C(\tau) \equiv& \frac{\sigma_{n}^{2}}{\sum_{m\neq 0}P_{m}(0)}
        \sum_{m\neq 0} \frac{P_{m+1}(0) - P_{m-1}(0)} {2 m}
        \label{constCoP} ,
    \end{align}
where the function $P_{m}(0)=\int dq\,dw P(m,q,w)$ is the probability distribution of the particle current, which is compatible with the Skellam distribution~\cite{Skellam1946}
(Appendix \ref{ssec:SD}),
    \begin{align}
        P_{m}(0) = PS(m,n^{+},n^{-}) \,,
        \;\;\;\;
        n^{\pm} \equiv \frac{\sigma^{2}_{n} \pm \langle n \rangle}{2} \, .
    \end{align}
Here this distribution corresponds to the bidirectional Poisson distribution for particle transport in the isothermal case $\delta \beta =0$ (Appendix \ref{ssec:BdPP}).
From Eqs. (\ref{aveCoP}) and (\ref{constCoP}), we observe that the coefficient $C(\tau)$ dominates the time dependence of the average COP since the cumulants (\ref{Averages}) and (\ref{Vars}) are proportional to $\tau$.

\subsubsection{Long-time limit \label{sssec:LTL}}
By taking $\tau \to \infty$, we derive ${\rm lim}_{\tau \to \infty}C(\tau)=0$ and confirm that the COP approaches that of the macroscopic system,
    \begin{align}
        \langle \phi \rangle_{\tau \to \infty} =& \frac{\langle q \rangle}{\delta \mu\, \langle n \rangle}
                                            \equiv \langle \phi \rangle_{\rm macro}
                                            \label{tauInfVal}.
    \end{align}
In addition, for the isothermal case $\delta \beta=0$, $\langle \phi \rangle_{\tau}$ is independent of $\tau$ because $C(\tau)=0$.
Thus, we always recover the COP of the macroscopic system (\ref{tauInfVal}), $\langle \phi \rangle_{\tau} = \langle \phi \rangle_{\rm macro}$.

In the long measurement time regime $\tau\gg 1$, we can evaluate the correction coefficient (\ref{constCoP}) approximately.
When the time becomes longer, the number of transferred particles increases, and the large fluctuations are suppressed.
Then the absolute value of the deviations $\delta q\equiv q-\langle q\rangle$ and $\delta n\equiv n-\langle n\rangle$ are smaller than these averages, $|\delta q| \ll |\langle q\rangle|$ and $ |\delta n| \ll |\langle n\rangle|$.
Under this condition, we can obtain the approximate formula~\cite{Crepieux2016} by expanding the $\langle q/n \rangle_{\tau}$ up to the second order in the deviations,
    \begin{align}
        \langle\phi\rangle_{\tau\gg1} =&
        \left< \frac{q}{w} \right>_{\tau\gg 1} = \frac{1}{\delta \mu}\left< \frac{\langle q \rangle + \delta q}{\langle n \rangle +\delta n} \right>_{\tau \gg 1} \nonumber \\
                        \simeq & \frac{\langle q \rangle}{\delta \mu \,\langle n \rangle}
                        +\frac{\sigma_{n}^{2}}{\delta \mu \,\langle n \rangle^{2}} \left( \frac{\langle q \rangle}{\langle n \rangle} -\frac{C_{nq}}{\sigma_{n}^{2}} \right).
                        \label{taulongtimeform}
    \end{align}

\subsubsection{Short-time limit \label{sssec:STL}}
In the short-time regime $\tau \ll 1$, the average COP (\ref{aveCoP}) can be evaluated by expanding $C(\tau)$ (\ref{constCoP}) in $\tau$ up to the first order, $C(\tau)\simeq -1+\langle n \rangle^{2}/4\sigma_{n}^{2}$.
Then we obtain the average COP as
    \begin{align}
        \langle \phi \rangle_{\tau\ll 1} \simeq& \frac{C_{nq}}{\delta \mu\, \sigma_{n}^{2}}
                + \frac{\langle n \rangle^{2}}{4\,\delta \mu\, \sigma_{n}^{2}} \left( \frac{\langle q \rangle}{\langle n \rangle} -\frac{C_{nq}}{\sigma^{2}_{n}} \right) .
                                            \label{tau1stOrder}
    \end{align}
The meaning of the approximation is as follows.
As the measurement time becomes shorter, the number of tunneling events decreases.
Eventually, at $\tau \to 0$, the joint probability distribution of the currents and Joule heat describes at most a single tunneling event (Appendix \ref{ssec:STRProb}).

Furthermore, we can derive ${\rm lim}_{\tau \to 0}C(\tau)=-1$ and obtain the COP average at $\tau \to 0$ as
    \begin{align}
		\langle \phi \rangle_{\tau \to 0} =& \frac{C_{nq}}{\delta \mu\, \sigma_{n}^{2}}
											\label{tau0Val}.
    \end{align}
Here we remark that this formula is compatible with the factor of proportionality of a linear minimum-mean-square-error estimator of the heat current for the particle current (Appendix \ref{ssec:Regression}).

\subsection{Fluctuation theorem\label{sec:FTCoP}}
In the general case, the fluctuation theorem~\cite{Esposito2009,Campisi2011,UtsumiPRB2014}
for the CGF of the number of particles and the amount of energy that enter lead $r$, $\tilde{W}(\{ \lambda_{r},\xi_{r}\}_{r=R,L})$, is written as
    \begin{align}
        \tilde{W}(\{ \lambda_{r},\xi_{r}\}_{r=R,L}) = \tilde{W}(\left\{ -\lambda_{r} + i\beta_{r} \mu_{r},-\xi_{r}+i\beta_{r} \right\}_{r=R,L}) \label{generalFT},
    \end{align}
where $\lambda_{r}$ and $\xi_{r}$ are the counting fields of the increases in the number of the particles and the amount of the energy in lead $r$, respectively.
They are defined in our convention as
    \begin{align}
        \lambda_{r} \equiv& \ s_{r}\frac{\lambda}{2} - \mu_{r} \left( \Xi + s_{r}\frac{\xi}{2} \right), \\
        \xi_{r} \equiv& \ \Xi + s_{r}\frac{\xi}{2},
    \end{align}
where $s_{r}$ is utilized in (\ref{CountField}).
When the measurement time is sufficiently long, which is always the case for a tunnel junction, the CGF depends only on the differences between the counting fields of the right and left leads,
    \begin{align}
        \tilde{W}(\{ \lambda_{r},\xi_{r}\}_{r=R,L}) \to \tilde{W}(\lambda_{R}-\lambda_{L} ,\xi_{R}-\xi_{L}).
    \end{align}
Thus, we can add an arbitrary constant in the transformation of $\lambda_{r}$ and $\xi_{r}$ in (\ref{generalFT}),
    \begin{align}
        \lambda_{r} \to& -\lambda_{r} - i\beta_{r}\mu_{r} +i \alpha \beta \bar{\mu},\\
        \xi_{r} \to& -\xi_{r} + i\beta_{r} -i\alpha\beta,
    \end{align}
where $\alpha$ is a constant that is not determined uniquely.
From the above equations, we derive the generalized form of the steady-state current fluctuation theorem,
    \begin{align}
        &W(\lambda,\xi,\Xi) \nonumber \\
        &= W (- \lambda + i \alpha \beta\,\delta\mu ,-\xi+i\delta\beta,-\Xi+ i(1-\alpha)\beta). \label{FR}
    \end{align}
Equation (\ref{FR}) indicates that the interchange between the thermodynamic forces of the particle current and the Joule heat does not change the physical meaning.
The thermodynamic force associated with the Joule heat vanishes when $\alpha=1$,
    \begin{align}
        W(\lambda,\xi,\Xi) = W (- \lambda + i \beta\,\delta\mu ,-\xi+i\delta\beta,-\Xi), \label{FRa1}
    \end{align}
and that associated with the particle current vanishes when $\alpha=0$,
    \begin{align}
        W(\lambda,\xi,\Xi) = W (- \lambda,-\xi+i\delta\beta,-\Xi+ i\beta). \label{FRa0}
    \end{align}
From Eq. (\ref{FRa1}) and by utilizing Jensen's inequality, we can show that the entropy production rate is nonnegative, that is, the second law of thermodynamics\cite{VerleyPRE2014},
    \begin{align}
        \partial_{\tau}S \equiv \frac{\langle n \rangle}{\tau} \beta \frac{\delta \mu}{2}
                + \frac{\langle q \rangle}{\tau} \frac{\delta \beta}{2} \geq 0 . \label{EPrate}
    \end{align}
This fact indicates that the COP of the macroscopic system cannot exceed the Carnot limit~(\ref{ineqCarnot}),
    \begin{align}
        \langle \phi \rangle_{\rm macro}
        \leq
        \phi_{\rm C}, \label{CarnotIneq}
    \end{align}
where Eq. (\ref{CarnotIneq}) is valid when $\delta \mu\, \langle n \rangle >0$.
According to the definition of the macroscopic COP (\ref{tauInfVal}), one may think that it is possible to exceed the Carnot limit by making $\delta \mu$ smaller.
However, for a sufficiently small $\delta \mu$, the average particle current $\langle n \rangle$ becomes negative because the thermoelectric current which flows in the opposite direction becomes dominant.

\section{Modified Gaussian Approximation \label{sec:MGA}}
We comment that the Gaussian approximation (Appendix~\ref{ssec:GA}), which uses the multivariate Gauss distribution as the probability distribution of the currents, does not provide the average COP at $\tau \to 0$.
In this limit, the probability distribution of the COP calculated from the Gauss distribution (\ref{GApdf}), which is defined as
    \begin{align}
        P(\phi) \equiv \iiint_{-\infty}^{\infty}dn \, dq \, dw \,P(n,q,w) \, \delta \left( \phi - \frac{q}{w} \right),
    \end{align}
approaches the Cauchy distribution~\cite{Polettini2015} (\ref{CauchyGpdf}), whose average does not exist.

This flaw occurs because a finite amount of heat $q$ is carried by zero Joule heat $w=\delta\mu\,n$ in the Gaussian approximation.
Thus, we amend this by excluding the zero point of the Joule heat,
    \begin{align}
        \int_{-\infty}^{\infty} dn \ \Rightarrow \  \lim_{\eta \to 0} \left( \int_{-\infty}^{-\eta} + \int_{\eta}^{\infty} \right) dn.
    \end{align}
This manipulation corresponds to cutting off the tail of the Cauchy distribution.
Thereby, we obtain the average COP as
    \begin{align}
        \langle\phi\rangle_{\tau} =& \frac{\langle q\rangle}{\delta\mu\,\langle n\rangle}
                                  + \frac{C_{\tau}^{GA}}{\delta\mu} \left( \frac{ \langle q \rangle }{ \langle n \rangle } -\frac{C_{nq}}{\sigma^{2}_{n}} \right),
                                  \label{DawsonAppr}\\
        C_{\tau}^{GA} \equiv& \left[ -1 + \frac{2\,\langle n\rangle}{\sqrt{2\sigma_{n}^{2}}} DF \left( \frac{\langle n\rangle}{\sqrt{2\sigma_{n}^{2}}} \right) \right],\label{DawsonApprC}
    \end{align}
where $DF(x)$ is a Dawson integral (Appendix~\ref{ssec:DF}).
The coefficient $C_{\tau}^{GA}$ determines the time dependence of the average COP.
The coefficient also has a minimum or maximum value at ${\langle n\rangle}/{\sqrt{2\sigma_{n}^{2}}}=1.5019\cdots$ (see Appendix~\ref{ssec:DF}).
Here the averages of the currents appearing in (\ref{DawsonAppr}) are given by (\ref{lincum}).
Within the Gaussian approximation, the variances of the particle and the heat currents and covariance are written as
    \begin{align}
        \sigma_{n}^{2}=&\tau L_{11}, \label{LinVarn}\\
        C_{nq}=&\tau L_{12}=\tau L_{21}, \label{LinCov} \\
        \sigma_{q}^{2}=&\tau L_{22}, \label{LinVarq}
    \end{align}
where $L_{ij}$ are the linear transport coefficients.

\section{Time Dependence of COP\label{sec:TECoP}}

\subsection{Evaluation of the statistical properties \label{ssec:Eval}}
We summarize that the average COP, which depends on the measurement time $\tau$, is generally written as
    \begin{align}
        \langle\phi\rangle_{\tau} =& \langle\phi\rangle_{\rm macro} +C(\tau) \,\delta\phi,
    \end{align}
where we define a time-independent quantity $\delta\phi$ as
    \begin{align}
        \delta\phi \equiv& \frac{1}{\delta\mu} \left( \frac{\langle q\rangle}{\langle n\rangle}-\frac{C_{nq}}{\sigma_{n}^{2}} \right).
    \end{align}
The approximated coefficients $C(\tau)$ are summarized in the following.
    \begin{align}
        C(\tau) \simeq& \ \frac{1}{2\tilde{n}^{2}}, \hspace{3.1em} {\rm (long \mathchar`- time\ limit}\  \tau\gg 1 {\rm )}\\
        C(\tau) \simeq& -1+\frac{\tilde{n}^{2}}{2}, \hspace{1em} {\rm (short \mathchar`- time\ limit}\  \tau\ll 1 {\rm )} \\
        C(\tau) \simeq& -DF'(\tilde{n}). \hspace{1em} {\rm (Gaussian\ approximation)} \label{GACtau}
    \end{align}
Here $DF'(x)$ is the derivative of the Dawson function and $\tilde{n} \equiv {\langle n\rangle}/{\sqrt{2\sigma_{n}^{2}}} \propto \sqrt{\tau}$.
In order to derive the average COP, we have to evaluate the cumulants (\ref{Averages}) and (\ref{Vars}).
Then it is necessary to calculate the conductivity $G(\xi)$ and the electromotive field $\epsilon(\xi)$ in (\ref{Averages}) [(\ref{N1}) and (\ref{N2})].
For this purpose, we perform (1) a linear approximation of the DOS and (2) a numerical calculation.

\subsubsection{Linear DOS approximation \label{sssec:LDOS}}
To perform the integration in (\ref{N1}) and (\ref{N2}), we expand the DOS $D_{R}(\epsilon)D_{L}(\epsilon)$ in terms of the energy $\epsilon - \bar{\mu}$ up to the first order,
    \begin{align}
        \frac{D_{R}(\epsilon)D_{L}(\epsilon)}{D_{R}(\bar{\mu})D_{L}(\bar{\mu})} \simeq 1+ \left( \epsilon -\bar{\mu} \right) D_{rat}. \label{DOSappr}
    \end{align}
Furthermore, we expand the Fermi-Dirac distribution in terms of $\delta\mu$ and $\delta\beta$ up to the leading order.
Thereby, we derive the analytic formulas of the conductivity (\ref{N3}) and the electromotive force (\ref{N4}).
We remark that the matrix of the linear transport coefficients (\ref{MtrxLR}) is obtained by picking out the linear parts in $\delta\mu$ and $\delta\beta$ from $G(\xi)$ and $\epsilon(\xi)$.

Here we discuss the validity of this approximation.
For the Gaussian approximation~\cite{Polettini2015} case, the validity is evaluated by checking the semi-positive definiteness of the linear transport matrix (\ref{MtrxLR}), $\det{\bf L} \geq 0$ and $L_{ii} \geq 0$.
Then the determinant of the linear transport matrix $\det{\bf L}=l_{\beta}(1-r^{2})$ corresponds to the the correlation coefficient between the particle and the heat currents, which is defined as
    \begin{align}
        r \equiv \frac{C_{nq}}{\sigma_{n} \sigma_{q}} \label{CorrCoeff}.
    \end{align}
Therefore, the condition
    \begin{align}
        r^{2}=
        \frac{C_{nq}^{2}}{\sigma_{q}^{2}\sigma_{n}^{2}} \leq 1 \label{sqCorrCoeff}
    \end{align}
is equivalent to $\det{\bf L} \geq 0$.
The proof of Eq.~(\ref{sqCorrCoeff}) for a quantum system is given in Appendix~\ref{ssec:CCUU}.

We summarize that in order to satisfy the physical consistency,
we must choose a parameter domain where the entropy production rate (\ref{EPrate}) is positive and the absolute value of the correlation coefficient (\ref{CorrCoeff}) is smaller than one.

\subsubsection{Numerical calculation \label{sssec:NDOS}}
We perform a numerical calculation of the cumulants (\ref{Averages}) and (\ref{Vars}).
In this case, we have to give a detailed DOS.
The results of the numerical calculation always satisfy the condition of the correlation coefficient (\ref{sqCorrCoeff}) (Appendix~\ref{ssec:CorrCBdP}).

In this section, we assume the simple DOS
    \begin{align}
        \frac{D_{R}(\epsilon)D_{L}(\epsilon)}{D_{R}(\bar{\mu})D_{L}(\bar{\mu})}
            = \left\{ \begin{array}{c}
                1+ \left( \epsilon -\bar{\mu} \right) D_{rat} \\ 0
            \end{array} \right.
            \begin{array}{c}
                 (\epsilon \geq \bar{\mu} - D_{rat}^{-1} ) \\ (\epsilon < \bar{\mu} - D_{rat}^{-1})
            \end{array}.
    \end{align}

\subsection{Results and Discussion \label{ssec:results}}
\begin{figure}
        \begin{center}
            \includegraphics[width=1.0 \columnwidth]{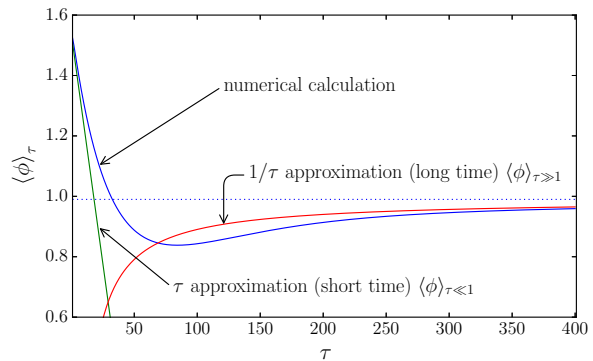}
        \end{center}
        \caption{
(Color online) Measurement time dependence of $\langle \phi \rangle_{\tau}$ and its approximate formulas $\langle \phi\rangle_{\tau\gg 1}$ (\ref{taulongtimeform}) and $\langle \phi\rangle_{\tau\ll 1}$ (\ref{tau1stOrder}).
The dotted line indicates the macroscopic value of the COP $\langle \phi \rangle_{\rm macro}$ (\ref{tauInfVal}).
The Carnot limit is $\phi_{\rm C}= 20.5$.
The entropy production rate is $\partial_{\tau}S=0.0581$.
Parameters: $D_{rat}/\beta=0.25$, $ \delta \beta/\beta =-0.0488 $, and $ \beta\, \delta \mu /2=0.256 $.
        }
        \label{fig:fcstimeevo}
    \end{figure}

In Fig.~\ref{fig:fcstimeevo}, we show the result of the numerical calculation and compare it with the long-time and short-time approximations (\ref{taulongtimeform}) and (\ref{tau1stOrder}).
We find that the average COP is not a monotonic function of the measurement time $\tau$.
The average COP first decreases, next reaches a minimum value, and finally approaches the macroscopic value $\langle \phi \rangle_{\rm macro}$.
This fact means that the initial average COP $\langle \phi\rangle_{\tau\ll 1}$ is larger than the macroscopic value $\langle \phi \rangle_{\rm macro}$ in the refrigerator.
The minimum is caused by the asymmetric shape of the probability distribution of the COP\cite{Polettini2015}.
Here, a downward convex curve appears under the conditions of the refrigerator, $\delta\beta<0$ and $\delta\mu>0$.
When the particle current flows from the hot lead to the cold lead, which is realized for $\delta\beta>0$ and $\delta\mu>0$, the curve is convex upward.

\begin{figure}
        \begin{center}
            \includegraphics[width=0.8 \columnwidth]{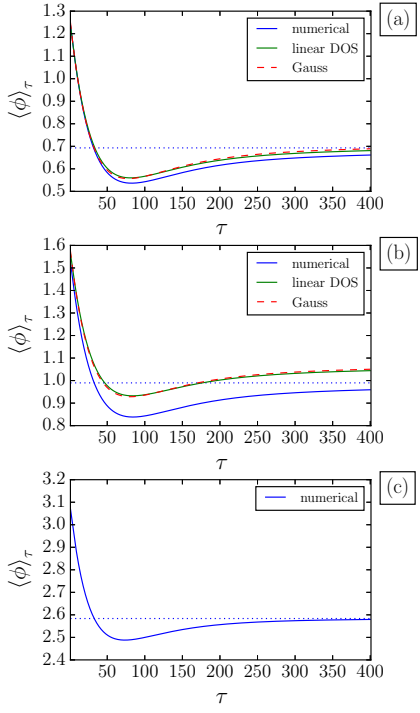}
        \end{center}
        \caption{
(Color online) Time dependence of $\langle \phi \rangle_{\tau}$ for (a) $D_{rat}/\beta=0.20$, (b) $D_{rat}/\beta=0.25$, and (c) $D_{rat}/\beta=0.75$.
The blue and green solid lines indicate the numerical calculation and the linear DOS approximation, respectively.
Dashed lines indicate the modified Gaussian approximation (\ref{DawsonAppr}).
The dotted lines indicate the convergence value $\langle \phi \rangle_{\rm macro}$ (\ref{tauInfVal}) for the numerical calculation.
The entropy production rates and the correlation coefficients for the numerical calculation are (a) $\partial_{\tau}S=0.0596$, $r=0.351$, (b) $\partial_{\tau}S=0.0581$, $r=0.425$, and (c) $\partial_{\tau}S=0.0574$, $r=0.717$.
Those for the linear DOS approximation are (a) $\partial_{\tau}S=0.0600$, $r=0.356$, (b) $\partial_{\tau}S=0.0579$, $r=0.449$, and (c) $\partial_{\tau}S=0.0369$, $r=1.41$.
Within the modified Gaussian approximation, (a) $\partial_{\tau}S=0.0594$, $r=0.363$, (b) $\partial_{\tau}S=0.0573$, $r=0.453$, and (c) $\partial_{\tau}S=0.0368$, $r=1.36$.
Parameters: $\delta\beta/2\beta = -0.0244$ and $\beta\,\delta\mu/2 = 0.256$.
The Carnot efficiency is $\phi_{C}=20.5$.
        }
        \label{fig:fcsvardos}
    \end{figure}

In Fig.~\ref{fig:fcsvardos}, we compare the results calculated by the modified Gaussian approximation (\ref{DawsonAppr}),
the linear DOS approximation (Eq. (\ref{aveCoP}) calculated using the approximate DOS (\ref{DOSappr}), and the numerical approach.
In this figure, we fix the affinities $\beta \,\delta\mu$ and $\delta\beta$ and vary the slope of the DOS $D_{rat}$.
Figure~\ref{fig:fcsvardos}(a) shows that these approximations have higher precision when the thermodynamic forces are sufficiently small and the DOS is almost flat.
In contrast, the approximations with a steeper DOS deviate from the numerical result (Fig.~\ref{fig:fcsvardos}(b)).
The approximate DOS (\ref{DOSappr}) can be negative, which reduces the accuracy of the approximations.
In Fig.~\ref{fig:fcsvardos}(c), in which the DOS has a much steeper slope, the linear DOS approximation results in an invalid correlation coefficient $r^{2}>1$, although the entropy production rate is positive.

\section{Summary\label{sec:SMR}}
We have investigated the effect of particle and heat current fluctuations on the COP of a nanoscale refrigerator.
In the framework of FCS, we obtained the joint CGF of particle and heat transfer in the bidirectional Poisson process within the second order perturbation expansion in the tunnel coupling.
On the basis of the resulting joint probability distribution, we derived the average COP, which turned out to be time-dependent and expressed with the Skellam distribution.
In the long-time limit, it approaches the macroscopic value.
We also improved the Gaussian approximation and obtained an average COP that is applicable at $\tau\to 0$ (Eqs. (\ref{DawsonAppr}) and (\ref{DawsonApprC})).

We numerically investigated the average COP.
We found that in the short-time regime, the average COP possesses a minimum.
Our numerical approach covers the parameter regime beyond the Gaussian approximation, which is limited to the case when the matrix of the linear response coefficients is positive-semidefinite.
We pointed out that in order to develop physically reasonable approximations in the nonequilibrium regime, we must check the correlation coefficient,
which is a measure of the linear dependence between the heat current and the particle current, in addition to the positivity of the entropy production rate.
In the tunnel regime, we proved that the absolute value of the correlation coefficient is always smaller than one (Appendix~\ref{ssec:CorrCBdP}).
However, by approximating the energy dependence of the DOS, the condition can be violated even though the entropy production rate is positive.
The physical consistency of the Gaussian approximation is evaluated by the positive-semidefiniteness of the linear transport matrix,
which ensures that the absolute value of the correlation coefficient is smaller than one and that the entropy production rate is positive.

\section*{Acknowledgments}

We thank Ora Entin-Wohlman and Amnon Aharony for their valuable discussions.
This work was supported by JSPS KAKENHI Grant Nos.~26400390 and~26220711.

\begin{appendix} \label{sec:APD}

\renewcommand{\thesubsection}{\thesection.\arabic{subsection}}
\renewcommand\thefigure{\thesection\arabic{figure}}

\section{First and Second Cumulants \label{ssec:cums}}

The cumulants $\langle n \rangle$, $\langle q \rangle$, $\sigma_{n}^{2}$, $\sigma_{q}^{2}$, and $C_{nq}$ are defined by using (\ref{CGF}),
    \begin{align}
        \langle n \rangle =&  \left. \partial_{i \lambda} W(\lambda,\xi,0) \right|_{\lambda=\xi=0}
        \, ,
        \label{n} \\
        \langle q \rangle =&  \left. \partial_{i \xi} W(\lambda,\xi,0) \right|_{\lambda=\xi=0}
        \, ,
        \label{q} \\
        \sigma_{n}^{2} \equiv& \langle (n-\langle n \rangle)^{2} \rangle \nonumber \\
               =&  \left. \partial_{i \lambda}^2 W(\lambda,\xi,0) \right|_{\lambda=\xi=0}
        \, ,\label{sigman} \\
        \sigma_{q}^{2} \equiv& \langle (q-\langle q \rangle)^{2} \rangle \nonumber \\
               =&   \left. \partial_{i \xi}^2 W(\lambda,\xi,0) \right|_{\lambda=\xi=0}
        \, ,\label{sigmaq} \\
        C_{nq} \equiv& \langle (n-\langle n \rangle)(q-\langle q \rangle) \rangle \nonumber \\
               =& \left. \partial_{i \lambda} \partial_{i \xi} W(\lambda,\xi,0) \right|_{\lambda=\xi=0}
        \, .
        \label{Cnq}
    \end{align}

\section{Calculation of the Characteristic Function\label{ssec:jPDF}}

By using the definition of the CGF (\ref{CGF}) and Eqs. (\ref{CGF1}) and (\ref{CGF2}), we derive the joint probability distribution as
    \begin{align}
        P(n,q,w) =&  \delta(w-n\delta \mu) \int_{-\infty}^{\infty} \frac{d \xi}{2\pi} P_{n}(\xi)e^{-iq\xi}
        , \label{Prob1} \\
        P_{n}(\xi) \equiv& \frac{\sum_{m=-\infty}^{\infty} I_{n-m}\left( G(\xi) \right)
        J_{m}\left( \varepsilon(\xi) \right)} {e^{F(0,0,0)- n\, \beta\, \delta \mu/2}}
        , \label{F_C}
    \end{align}
where $J_{n}(x)$ is the Bessel function and
    \begin{align}
        G(\xi) \equiv& \int_{-\infty}^{\infty}d\epsilon \,h(\epsilon)\, {\rm cosh}
        \left[\left( i\xi+\frac{\delta \beta}{2} \right)(\epsilon-\bar{\mu}) \right]
        \label{N1} \, , \\
        \varepsilon(\xi) \equiv&  \int_{-\infty}^{\infty}d\epsilon \,h(\epsilon)\, {\rm sinh}
        \left[\left( i\xi+\frac{\delta \beta}{2} \right)(\epsilon-\bar{\mu}) \right] \label{N2}
        \, .
    \end{align}
Here
    \begin{align}
        h(\epsilon) \equiv&
        \frac{\beta}{2} \, \frac{D_{R}(\epsilon)D_{L}(\epsilon)}{D_{R}(\bar{\mu})D_{L}(\bar{\mu})}
        \, \frac{ f_{L}(\epsilon)-f_{R}(\epsilon)}
                {{\rm sinh}\left( \frac{\delta \beta}{2}(\epsilon-\bar{\mu})+\beta \frac{\delta \mu}{2}\right)} \label{WF1}
        \, , \\
        f_{r}(\epsilon) =& \frac{1}{e^{\beta_{r}(\epsilon-\mu_{r})}+1} .
    \end{align}
By substituting Eq.~(\ref{Prob1}) into Eq.~(\ref{Prob2}), we derive the characteristic function of the probability distribution of the COP as
    \begin{align}
        \chi(\gamma) =\int d\phi P(\phi)e^{i\gamma \phi} =& \sum_{m\neq 0} P_{m}
        \left( \frac{i\gamma}{m\,\delta \mu} \right)
        .
    \end{align}
From the first derivative, we obtain the average COP (\ref{aveCoP}) as
    \begin{align}
        \langle \phi \rangle_{\tau} = \left.
                \partial_{i \gamma}
                \ln
                \chi(\gamma)
                \right|_{\gamma=0}
                \, .
    \end{align}

\section{Integration Performed around Equilibrium\label{ssec:IaE}}
The integral in Eqs.~(\ref{N1}) and (\ref{N2}) can be performed easily by approximating the weight function (\ref{WF1}) around the equilibrium.
We expand $D_R(\epsilon)D_{L}(\epsilon)$ around the average chemical potential $\bar{\mu}$ (\ref{DOSappr}) and expand the Fermi distribution function in $\beta\, \delta \mu$ and $\delta \beta /\beta$ in the leading order as
    \begin{align}
        h(\epsilon) \simeq&
        \beta \left( 1 + D_{rat}(\epsilon-\bar{\mu})\right)
        \frac{e^{\beta(\epsilon-\bar{\mu})}} {\left(e^{\beta(\epsilon-\bar{\mu})} +1 \right)^{2}}
        .\label{WF2}
    \end{align}
From the approximated weight function, we can calculate (\ref{N1}) and (\ref{N2}) by expanding $\cosh$ and $\sinh$ in $(i\xi+\delta \beta/2)(\epsilon-\bar{\mu})$ and using the methods explained in Appendix~\ref{ssec:IFD},
    \begin{align}
        G(\xi) =& \sum_{n=0}^{\infty}
                \frac{\left[ \frac{2 \pi i}{\beta}(i\xi+\frac{\delta \beta}{2}) \right]^{2n}}{2n!}
                B_{2n}\left( \frac{1}{2} \right)\\
                =& \left[ j_{0}\left( \pi \frac{i\xi+\frac{\delta \beta}{2}}{\beta}\right)\right]^{-1}
        ,\label{N3}
    \end{align}
and
    \begin{align}
        \varepsilon(\xi) =& D_{rat} \frac{\partial}{\partial (i\xi)} G(\xi)
                        = \frac{\pi}{\beta} D_{rat}
                        \frac{j_{1}\left( \pi \frac{i\xi+\frac{\delta \beta}{2}}{\beta} \right)}
                        {j_{0}\left(\pi \frac{i\xi+\frac{\delta \beta}{2}}{\beta} \right)^{2}} \,, \label{N4}
    \end{align}
where $B_{n}(x)$ is the Bernoulli polynomial and $j_{m}(z)$ is the spherical Bessel function of the $m$th order.
In the above calculations, we use the properties of the Bernoulli polynomial\cite{Olver2010},
    \begin{align}
        \frac{t e^{xt}}{e^{t}-1} &= \sum_{n=0}^{\infty} B_{n}(x)\frac{t^{n}}{n!}
        \,,
        \;\;\;\;
        (|t|<2\pi) \,,
        \\
        B_{2n+1}\left(\frac{1}{2} \right) &= 0 \,.
    \end{align}
Here, these results are valid for $|\pi(i\xi+\frac{\delta \beta}{2})/\beta|<\pi$;
however, we do not need to consider this condition because we set $i\xi =0$ when we calculate the cumulants, and the condition ${\delta \beta}/{2\beta}<1$ is always satisfied.

\section{Skellam Distribution \label{ssec:SD}}

The Skellam process is represented by the difference between two independent stochastic variables that obey the Poisson distribution\cite{Skellam1946}.
The Skellam distribution function is given by
    \begin{align}
        PS(n,c_{1},c_{2}) = e^{-\frac{c_{1}+c_{2}}{2}} \left( \sqrt{ \frac{c_{1}}{c_{2}} } \right)^{n} I_{n}\left( \sqrt{c_{1}c_{2}} \right) .
        \label{Skellamd}
    \end{align}
The characteristic function is
    \begin{align}
    \chi_{PS}(\lambda) = {\rm Exp}\left( -\frac{c_{1}+c_{2}}{2} + \frac{c_{1}e^{i \lambda} +c_{2} e^{-i \lambda}}{2} \right)
    \ \ .
    \end{align}
The average and variance are
    \begin{align}
        \bar{n} =& c_{1}-c_{2}, \\
        \sigma_{n}^{2} =& c_{1} + c_{2} \ \ .
    \end{align}

\section{Bidirectional Poisson Distribution \label{ssec:BdPP}}

When the temperatures of the left and right leads are the same ($\delta \beta=0$) and we ignore the counting fields $\xi$ and $\Xi$, the non-normalized CGF~(\ref{CGF2}) becomes
    \begin{align}
        F(\lambda) =& \tau g \frac{{\rm cosh}\left( \beta\, \delta \mu/ 2 + i\lambda \right)} {{\rm sinh}\left( \beta\, \delta \mu/2 \right)},
        \\
        g =& \int_{-\infty}^{\infty} d\epsilon \, d_{\beta}(\epsilon) \,\left( f_{L}(\epsilon)-f_{R}(\epsilon) \right),\label{Cnd}
        \end{align}
where
    \begin{align}
        d_{\beta} (\epsilon) \equiv \frac{\beta}{2} \frac{D_{L}(\epsilon)D_{R}(\epsilon)}{D_{L}(\bar{\mu})D_{R}(\bar{\mu})}.
    \end{align}
From the inverse Fourier transform, we obtain the bidirectional Poisson distribution [Eq. (E4) in Ref.~\onlinecite{Esposito2009}] for the particle transmission as
    \begin{align}
        P(n) =& \frac{e^{n \beta\, \delta \mu/2} } {e^{ \tau g \coth \left(\beta\, \delta \mu/2 \right)} }
        I_{n} \left( \frac{\tau g}{{\rm sinh} \left(\beta\, \delta \mu/2 \right)} \right) ,
    \end{align}
where $I_{n}(x)$ is the modified Bessel function.
The bidirectional Poisson distribution is represented by the Skellam distribution (\ref{Skellamd}) as
$P(n)=PS(n,c_{1},c_{2})$, where
    \begin{align}
        c_{1} =& 2\tau\int_{-\infty}^{\infty} d\epsilon \, d_{\beta}(\epsilon) \, f_{R}(\epsilon)(1-f_{L}(\epsilon))
         = \frac{2\tau g}{1-e^{-\beta\, \delta \mu}} \, ,
        \\
        c_{2} =& 2\tau\int_{-\infty}^{\infty} d\epsilon \, d_{\beta}(\epsilon) \, f_{L}(\epsilon)(1-f_{R}(\epsilon))
         = \frac{2\tau g}{e^{\beta\, \delta \mu}-1} \, .
    \end{align}

The average and variance are derived as
    \begin{align}
        \langle n \rangle =& \tau g \\
        \sigma_{n}^{2} =& \langle n \rangle \, {\rm coth}\left( \beta \frac{\delta \mu}{2} \right) .
    \end{align}
When the DOS is sufficiently flat around $\bar{\mu}$, namely $D_{L}(\epsilon)D_{R}(\epsilon) \simeq D_{L}(\bar{\mu})D_{R}(\bar{\mu})$, (\ref{Cnd}) is approximated as $g\simeq\beta \, \delta\mu/2$.

\section{Joint Probability Distribution in Short-Time Regime \label{ssec:STRProb}}
In the short time regime $\tau \ll 1$, the CGF (\ref{CGF1}) is proportional to the measurement time $\tau$.
The joint probability distribution is calculated as
    \begin{align}
        P(n,q,w)=& \int \frac{d\lambda \,d\xi\, d\Xi}{(2\pi)^{3}}\, e^{W(\lambda,\xi,\Xi)-in\lambda-iq\xi-iw\Xi} \nonumber \\
        \simeq& \int \frac{d\lambda \,d\xi\, d\Xi}{(2\pi)^{3}}\, e^{-in\lambda-iq\xi-iw\Xi} \left[1 + W(\lambda,\xi,\Xi) \right] \nonumber \\
        =& \left( 1-F(0,0,0)\, \right) \, \delta_{n,0} \, \delta(q) \, \delta(w) \nonumber \\
        &+ \frac{\tau}{2}\,e^{(\delta \beta \,q + \beta \delta \mu)/2}\, h(\bar{\mu}+q) \, \delta_{n,1} \, \delta(w-\delta \mu) \nonumber \\
        &+ \frac{\tau}{2}\,e^{(\delta \beta \,q - \beta \delta \mu)/2}\, h(\bar{\mu}-q) \, \delta_{n,-1} \, \delta(w+\delta \mu). \label{OneEvent}
    \end{align}
The terms appearing in the third, fourth, and fifth lines of (\ref{OneEvent}) represent the probabilities that a particle does not flow,
that a particle flows from the left lead to the right lead, and that a particle flows in the opposite direction, respectively.
Thus, we can regard the joint probability distribution as having the reduced support $n \in \left\{ -1,0,1 \right\}$.

\section{Linear Minimum-Mean-Square-Error (MMSE) Estimator\label{ssec:Regression}}
When the joint probability distribution of $n$ and $q$ is given as $P(n,q)$, the linear estimator of $q$ for $n$ and the mean square error are written as
    \begin{align}
        \eta(n) =& a_{1} n + a_{2} , \\
        {\rm MSE} =& \int dn\,dq \left( q-\eta(n) \right)^{2}P(n,q).
    \end{align}
The coefficients $a_{1}$ and $a_{2}$ can be derived by using the linear minimum mean square error (MMSE).
The linear MMSE estimator is written as
    \begin{align}
        \eta(n)   =& \, \frac{C_{nq}}{\sigma^{2}_{n}} \, (n - \langle n \rangle)+ \langle q \rangle,
    \end{align}
where $C_{nq}$ and $\sigma^{2}_{n}$ are the covariance between $n$ and $q$ and the variance of $n$, respectively.
Furthermore, we can rewrite the estimator as
    \begin{align}
        \eta(n) =& \, \frac{C_{nq}}{\sigma^{2}_{n}} \, n + \langle n \rangle \left[ \frac{\langle q \rangle}{\langle n \rangle} - \frac{C_{nq}}{\sigma^{2}_{n}} \right], \label{mmseestimator}
    \end{align}
and confirm that the rate $\eta(n)/n$ approaches constant $C_{nq}/\sigma_{n}^{2}$ when $\langle n \rangle \to 0$.

Although the above estimator is obtained for a continuous probability distribution, this procedure is also valid for a discrete distribution.

\section{Gaussian Approximation\label{ssec:GA}}
Here, we summarize the Gaussian approximation\cite{Polettini2015}.
By expanding (\ref{CGF2}) in $\delta \mu$, $\delta \beta$, $\Xi$ , $i\xi$, and $i\lambda$ up to the second order, we obtain the CGF and the joint probability distribution as
    \begin{align}
        W(\lambda,\xi,\Xi) =& \frac{\tau}{2}
                              \left( \begin{array}{c}
                                        i\lambda+i\Xi \,\delta\mu \\ i\xi
                                     \end{array}
                              \right)^{T}
                              {\bf L}\,
                              \left( \begin{array}{c}
                                        i\lambda+i\Xi \,\delta\mu +\beta \delta\mu \\ i\xi +\delta\beta
                                     \end{array}
                              \right) \label{GACGF} \\
        P(n,q,w) =& \frac{\delta(w-n\,\delta \mu)}{4\pi \tau \sqrt{{\rm det} \, {\bf L}}}
        e^{-({\bf j-\langle j \rangle})^{\rm T} {\bf L}^{-1}({\bf j-\langle j \rangle})/(2 \tau)}
        , \label{GApdf} \\
        {\bf j} =& \left( \begin{array}{c}
                                n \\ q
                          \end{array} \right) \,, \;\;\;\;
        \langle{\bf j}\rangle = \left(
            \begin{array}{c}
                \langle n \rangle \\ \langle q \rangle
            \end{array}
        \right) .
    \end{align}
By using this joint probability distribution (\ref{GApdf}), we obtain the probability distribution of the COP,
    \begin{align}
        P(\phi) =& \frac{\delta \mu\, e^{-\frac{\tau}{2}{\bf \langle j \rangle^{\rm T}}{\bf L}^{-1}{\bf \langle j \rangle}}}
                        {\pi a(\phi) \sqrt{{\rm det}{\bf L}}}
                    \left[ 1+ \sqrt{\pi \tau} h(\phi) e^{\tau h(\phi)^{2}}{\rm erf}(\sqrt{\tau}h(\phi)) \right]
                ,\label{CoPpdf} \\
        a(\phi) =& \frac{(\delta \mu\,\phi)^{2}-2D_{rat}l_{\beta}\,\delta \mu\, \phi +l_{\beta}}{{\rm det}{\bf L}} ,\\
        h(\phi) =& \frac{\frac{\delta \mu}{2} (\beta + \delta \beta \,\phi)}{\sqrt{2a(\phi)}} ,
    \end{align}
where ${\rm erf}(x)$ is the error function.
The probability distribution function approaches the Cauchy distribution at $\tau\to0$,
    \begin{align}
        P(\phi) \to \frac{1}{\pi}\frac{\delta\mu\,\sqrt{\det{\bf L}}}{(\delta\mu\,\phi-D_{rat}l_{\beta})^{2}+l_{\beta}-l_{\beta}^{2}D_{rat}^{2}} \label{CauchyGpdf};
    \end{align}
therefore, the average COP cannot be obtained since the average of the Cauchy distribution is not defined.

At ${\rm det} \, {\bf L}=0$, the ``tight-coupling case''~\cite{Polettini2015, KedemTFC1965} is realized.
Then we obtain the delta distribution from the Gaussian approximation,
    \begin{align}
        P(n,q ,w) = \frac{e^{-\frac{1}{2 \tau}(n-\langle n \rangle)^{2}}}{\sqrt{2 \pi \tau}}
                    \delta \left( q-\sqrt{l_{\beta}} n \right) \, \delta(w-n\,\delta \mu) , \label{TCGauss}
    \end{align}
which immediately leads to the probability distribution of the COP~(\ref{TightCoup}).

\section{Dawson Function \label{ssec:DF}}
The Dawson function is defined as
    \begin{align}
        DF(x)\equiv& e^{-x^{2}} \int_{0}^{x} dt \, e^{t^{2}}\\
            =& \frac{1}{2\sqrt{\pi}} \lim_{\eta\to 0}\left( \int_{-\infty}^{x-\eta} + \int_{x+\eta}^{\infty} \right) dt
                \frac{e^{-t^{2}}}{x-t}.
    \end{align}
We can also confirm that the Dawson function satisfies the following differential equation:
    \begin{align}
        \frac{d}{dx}DF(x) = -2xDF(x) +1.
    \end{align}
This derivative $(d/dx)DF(x)$, which appears in Eq. (\ref{GACtau}), has extrema at the points where the condition
    \begin{align}
        DF(x) = \frac{x}{2x^{2}-1} \label{convexDDF}
    \end{align}
is satisfied ($x_{c}^{\pm}=\pm1.5019\cdots$, $DF(x_{c}^{\pm})=\pm0.42768\cdots$).

\section{Correlation Coefficient in Quantum System\label{ssec:CCUU}}
Here we prove Eq. (\ref{sqCorrCoeff}) generally.
The Schr\"{o}dinger inequality\cite{Sakurai1985, Watanabe2014} is written as
    \begin{align}
        \sqrt{ \langle \delta\hat{I}^{2} \rangle \langle \delta\hat{J}^{2} \rangle } \geq \frac{1}{2} \sqrt{ \left| \langle [ \delta\hat{I}, \delta\hat{J} ] \rangle \right|^{2}
                                                                + \left| \langle \{ \delta\hat{I}, \delta\hat{J} \} \rangle \right|^{2} },
    \end{align}
where $\delta\hat{A}$ is the deviation of the operator $\hat{A}$, which is defined as $\hat{A}-\langle \hat{A} \rangle$.
Therefore, the correlation coefficient $r$ has the bounds
    \begin{align}
        |r| = \frac{\langle \{ \delta\hat{I}\delta\hat{J} \} \rangle}{2 \sqrt{ \langle \delta\hat{I}^{2} \rangle \langle \delta\hat{J}^{2} \rangle }}
          \leq & \frac{\langle \{\delta\hat{I}\delta\hat{J} \} \rangle}{\sqrt{\left| \langle [ \delta\hat{I}, \delta\hat{J} ] \rangle \right|^{2} + \left| \langle \{ \delta\hat{I}, \delta\hat{J} \} \rangle \right|^{2} }}.
    \end{align}
Here, $\sigma_{n}^{2} = \langle \delta\hat{I}^{2} \rangle$, $C_{nq} = \langle \{ \delta\hat{I}\,\delta\hat{J} \} \rangle/2$, and $\sigma_{q}^{2} = \langle \delta\hat{J}^{2} \rangle$.
Generally, in a quantum system, the upper bound of the correlation coefficient is smaller than one.
The upper bound is one when the operators $\delta\hat{I}$ and $\delta\hat{J}$ are commutative.

\section{Correlation Coefficient\label{ssec:CorrC}}
\setcounter{figure}{0}
\subsection{Correlation coefficient of the bidirectional Poisson process \label{ssec:CorrCBdP}}
In the bidirectional Poisson case, by using (\ref{N1}), (\ref{N2}), (\ref{Averages}), and (\ref{Vars}), Eq. (\ref{sqCorrCoeff}) is written as
    \begin{align}
        \frac{\int_{-\infty}^{\infty} \, d\epsilon \, \tilde{h}(\epsilon) \, (\epsilon - \bar{\mu})^{2}}{\int_{-\infty}^{\infty}d\epsilon \, \tilde{h}(\epsilon)}
        - \left[ \frac{\int_{-\infty}^{\infty}d\epsilon \, \tilde{h}(\epsilon) \, (\epsilon - \bar{\mu})}{\int_{-\infty}^{\infty}d\epsilon \, \tilde{h}(\epsilon)} \right]^{2}
        \geq 0,\label{WFineq}
    \end{align}
where
    \begin{align}
        \tilde{h}(\epsilon) \equiv h(\epsilon) \,{\rm cosh} \left( \frac{\delta\beta}{2}(\epsilon-\bar{\mu})+\beta \frac{\delta\mu}{2} \right).
    \end{align}
The weight function $h(\epsilon)$ defined in Eq.~(\ref{WF1}) is non-negative, hence $\tilde{h}(\epsilon)$ is non-negative in all domains in $\epsilon$.
Therefore, the right-hand side of (\ref{WFineq}) can be interpreted as the variance of the probability density function $\tilde{h}(\epsilon)/\int d\epsilon\tilde{h}(\epsilon)$.
Then the inequality (\ref{WFineq}) is always satisfied, namely, the square of the correlation coefficient is smaller than one in the entire parameter domain.

When the heat and the particle currents are perfectly correlated, namely $|r|=1$, the tight-coupling case~\cite{Polettini2015, KedemTFC1965} is realized.
Then the tunnel junction acts as an energy filter, therefore the COP becomes time-independent.
The tight-coupling case is realized when the density function $\tilde{h}(\epsilon)/\int d\epsilon\tilde{h}(\epsilon)$ is proportional to the delta function.

\subsection{Correlation coefficient within the linear DOS approximation}
The modified Gaussian approximation (\ref{DawsonAppr}) and the linear DOS approximation (\ref{DOSappr}) or (\ref{WF2}) can break the physical consistency.
Within these approximations, the weight function $\tilde{h}(\epsilon)$ can be negative.
Thus, the condition (\ref{sqCorrCoeff}) can be violated by varying the parameter $D_{rat}$ in the approximate DOS (\ref{DOSappr}) or (\ref{WF2}).

\begin{figure}
        \begin{center}
            \includegraphics[width=0.8 \columnwidth]{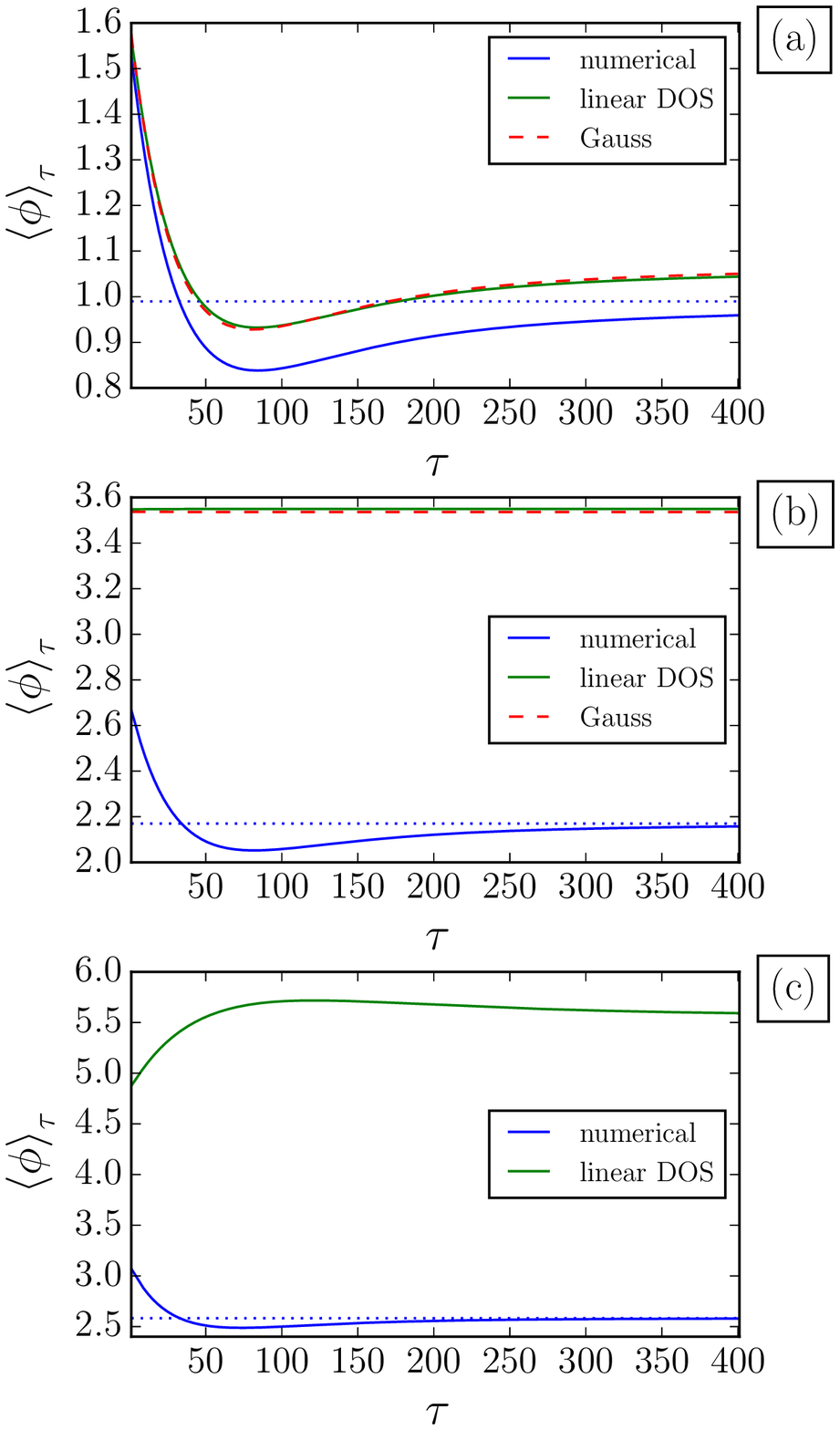}
        \end{center}
        \caption{
(Color online) Time dependence of $\langle \phi \rangle_{\tau}$ for (a) $D_{rat}/\beta=0.25$,
(b) $D_{rat}/{\beta}=\sqrt{3/\pi^{2}}= 0.551$ (tight-coupling condition),
and (c) $D_{rat}/\beta=0.75$.
The blue and green solid lines indicate the results obtained by the numerical calculation and the linear DOS approximation, respectively.
Dashed lines are the results obtained by the modified Gaussian approximation (\ref{DawsonAppr}).
The Carnot limit is $\phi_{\rm C}= 20.5$.
The entropy production rates in the linear DOS approximation are (a) $\partial_{\tau}S=0.0579$, (b) $\partial_{\tau}S=0.0453$, and (c) $\partial_{\tau}S=0.0369$.
Parameters: $ \delta \beta/\beta =-0.0488 $ and $ \beta\, \delta \mu /2=0.256 $.
        }
        \label{fig:plotfcs}
    \end{figure}

In Fig.~\ref{fig:plotfcs}, we show the three cases, (a) $\det {\bf L}>0$, (b) $\det {\bf L}=0$, and (c) $\det {\bf L}<0$.
In Fig.~\ref{fig:plotfcs}(a), the Gaussian approximation and the linear DOS approximation are valid since $\det {\bf L}>0$.
Figure~\ref{fig:plotfcs}(b) shows the time dependence when $\det{\bf L}=0$.
Since
    \begin{align}
        \det{\bf L} = l_{\beta}(1-r^{2}) =0,
    \end{align}
the tight-coupling condition~\cite{Polettini2015, KedemTFC1965} is realized and the joint probability distribution within the Gaussian approximation becomes a delta distribution (\ref{TCGauss}).
Thus, we obtain
    \begin{align}
        P(\phi) = \delta \left( \phi - \frac{1}{\beta\, \delta \mu}\sqrt{\frac{\pi^{2}}{3}}\right).
        \label{TightCoup}
    \end{align}
This results in a time-independent average COP, $\langle \phi \rangle_{\tau}=(\sqrt{\pi^{2}/3})/{\beta\, \delta \mu}$.
On the other hand, in the linear DOS approximation, the correlation coefficient exceeds one, $r=1.02$, therefore it is not physically consistent.
Here we comment on (\ref{TightCoup}).
The correlation between the left and right leads increases after the tunneling Hamiltonian is introduced.
Therefore, the cumulants of the particle and the heat current are time-dependent.
However, in the tight-coupling case, the heat and the particle currents $q$ and $n$ are perfectly linearly correlated, $q = \sqrt{l_\beta} n$, which is mentioned in (\ref{TCGauss}).
Thus, the COP, which is the ratio between $q$ and $n$, is time-independent, although $q$ and $n$ are time-dependent.

In Fig.~\ref{fig:plotfcs}(c), we show the time dependence when ${\rm det} \,{\bf L}<0$.
The Gaussian approximation and the linear DOS approximation are no longer applicable, although the entropy production rate is positive, $\partial_{\tau}S>0$.
This is because the correlation coefficients of the Gaussian approximation and the linear DOS approximation are $r=1.36$ and $r=1.41$, respectively, and then the condition (\ref{sqCorrCoeff}) is violated.
In contrast, the result of the numerical calculation still satisfies the condition, $-1<r=0.717<1$.

\section{Logistic Distribution and Weight Integration\label{ssec:IFD}}
\setcounter{figure}{0}
In our calculations, we frequently encounter integrals weighted by the logistic distribution.
We define the weighted integral as
    \begin{align}
        S_{u}[(\epsilon - m)^{n}]\equiv&\int_{-\infty}^{\infty} d\epsilon
        \frac{e^{u(\epsilon-m)}}{\left( e^{u(\epsilon -m)} + 1 \right)^{2}} \left( \epsilon-m \right)^{n}\\
        =& \frac{1}{u^{n+1}}\int_{0}^{\infty} dx \frac{({\rm ln}x)^{n}}{(x+1)^{2}} \ \ ,
    \end{align}
where $m$ and $b \neq 0$ are the parameters of the logistic distribution.
Then, the integral
    \begin{align}
        s_{n} \equiv& \int_{0}^{\infty} dx
        \frac{({\rm ln}x)^{n}}{(x+1)^{2}}  \label{lnx}
    \end{align}
can be performed along the contour in the complex plane depicted in Fig.~\ref{fig:Complex}.
As a result, we obtain the recurrence relation
    \begin{align}
        1 = \frac{2^{n-1}}{n}\sum_{k=0}^{n-1} \left(\begin{array}{c} n \\ k \end{array} \right) \frac{s_{k}}{(2 \pi i)^{k}} \ \ .
    \end{align}
The recurrence relation is identical to that of the Bernoulli polynomials\cite{Olver2010}
    \begin{align}
        n\,x^{n-1} = \sum_{k=0}^{n-1}
                \left(\begin{array}{c} n \\ k \end{array} \right)
                B_{k}(x) \ \ . \label{ber}
    \end{align}
By comparing these relations, we obtain
    \begin{align}
        s_{n} = \left( 2 \pi i \right)^{n} B_{n}\left( \frac{1}{2} \right)
        \, .
    \end{align}
Finally, we obtain the solution
    \begin{align}
        S_{u}[(\epsilon - m)^{n}] = \frac{1}{u} \left( \frac{2 \pi i}{u} \right)^{n}
                                         B_{n}\left( \frac{1}{2} \right) \, .
    \end{align}

    \begin{figure}[ht]
        \begin{center}
            \includegraphics[width=0.7 \columnwidth]{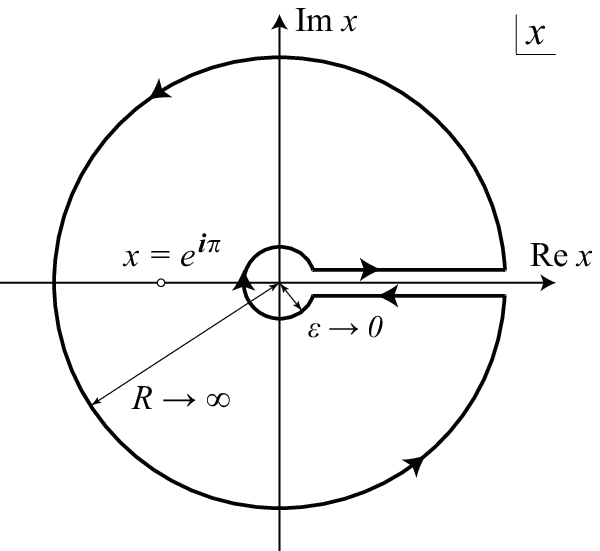}
        \end{center}
        \caption{
(Color online) Contour in the complex plane.
        }
        \label{fig:Complex}
    \end{figure}
\end{appendix}

\end{document}